\documentclass[runningheads]{llncs}

\usepackage{eccv}

\usepackage{eccvabbrv}
\usepackage[table]{xcolor}
\definecolor{cvprblue}{rgb}{0.21,0.49,0.74}
\definecolor{rowgray}{RGB}{242,242,242}
\definecolor{ourred}{RGB}{199,106,122}
\definecolor{ForestGreen}{RGB}{34,139,34}
\definecolor{BrickRed}{RGB}{178,34,34}
\usepackage{booktabs,multirow,makecell}
\usepackage{pifont,xcolor}
\usepackage{wrapfig}
\usepackage{amsmath}
\usepackage{amssymb}
\usepackage{algorithm}
\usepackage{diagbox}

\usepackage[noend]{algpseudocode}

\algnewcommand\algorithmicinput{\textbf{Input:}}
\algnewcommand\Input{\item[\algorithmicinput]}
\algnewcommand\algorithmicoutput{\textbf{Output:}}
\algnewcommand\Output{\item[\algorithmicoutput]}
\algnewcommand\algorithmicinit{\textbf{Initialize:}}
\algnewcommand\Init{\item[\algorithmicinit]}

\newcommand{\cmark}{\textcolor{ForestGreen}{\ding{51}}}
\newcommand{\xmark}{\textcolor{BrickRed}{\ding{55}}}

\newcommand{\best}[1]{\textbf{#1}}
\newcommand{\second}[1]{\underline{#1}}

\usepackage{graphicx}
\usepackage{booktabs}

\usepackage{hyperref}

\usepackage{orcidlink}

\begin{document}

\title{FlexiBrain: Resolution-Agnostic Voxel-Level Encoding for Native fMRI}

\titlerunning{FlexiBrain}

\author{Mo Wang\inst{1,2}\textsuperscript{\dag} \and
Wenhao Ye\inst{1}\textsuperscript{\dag} \and
Junfeng Xia\inst{1}\textsuperscript{\dag} \and
Minghao Xu\inst{2} \and
Hongkai Wen\inst{2}\textsuperscript{*} \and
Quanying Liu\inst{1}\textsuperscript{*}}

\authorrunning{M.~Wang et al.}

\institute{Southern University of Science and Technology \and
University of Warwick\\
\textsuperscript{\dag}Equal contribution. \textsuperscript{*}Corresponding authors.}

\maketitle

\begin{abstract}
The success of large-scale deep learning models in neuroscience is fundamentally constrained by severe data heterogeneity. Native fMRI data aggregated from diverse sources exhibit substantial variation in both spatial and temporal resolutions.
Consequently, most existing frameworks rely on lengthy, rigid preprocessing pipelines that enforce uniformity across datasets.
This practice introduces two critical limitations: (1) potential degradation of subject-specific anatomical information;
(2) significant computational overhead, often requiring hours of processing per subject.
Here, we propose FlexiBrain, a resolution-agnostic voxel-level encoding framework for native fMRI based on Mamba-JEPA. FlexiBrain defines patch sizes in real-world physical units and employs a dynamic patch resizing, thereby bypassing destructive spatial standardization while enabling direct ingestion of data in native space. We instantiate the framework using an efficient Mamba-JEPA backbone to model high-dimensional 4D fMRI signals. Across five diverse downstream neuroscience tasks, FlexiBrain consistently outperforms recent state-of-the-art methods, achieving gains of up to 12 percentage points without external data augmentation. Importantly, FlexiBrain functions as a seamless plug-in module, substantially reducing preprocessing costs and accelerating the development of robust voxel-level fMRI foundation models. Code is available at \url{https://github.com/OneMore1/FlexiBrain}.

\end{abstract}

\section{Introduction}
\label{sec:intro}

\begin{figure}[t]
  \centering
   \includegraphics[width=1\linewidth]{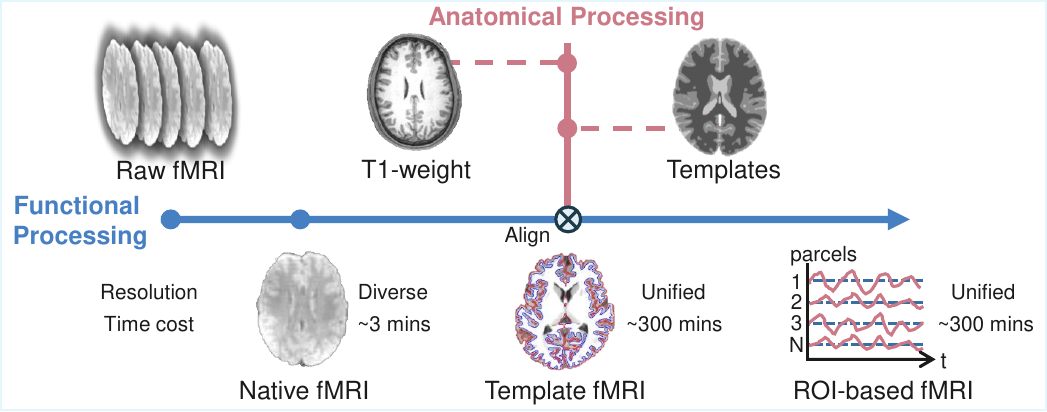}
   \caption{\textbf{fMRI preprocessing.}
   \textbf{Native fMRI} is rapid and retains subject-specific anatomical information. However, it presents a challenge for deep learning due to its highly diverse resolution and non-aligned spatial coordinates.
   \textbf{Template fMRI} requires resource-intensive  registration to align the image to a standard template brain, often consuming hours per subject. This eliminates individual differences, resulting in a unified and standardized resolution.
   \textbf{ROI-based fMRI} is derived from the registered Template fMRI using brain parcellation. It is reduced to a 2D representation (channels $\times$ time), summarizing activity across defined regions of interest.
   }
   \label{fig:motivation}
\end{figure}

Functional Magnetic Resonance Imaging (fMRI) is an indispensable tool in modern neuroscience, providing crucial, non-invasive insights into human brain activity. By providing voxel-level images of brain activity, it has fundamentally advanced our understanding of cognitive processes and has been instrumental in diagnosing and tracking numerous neurological disorders since its inception in the 1990s~\cite{belliveau1991functional}. Although early scanning protocols were often constrained by low spatial and temporal resolutions, recent technological advancements have yielded data with vastly higher resolution and developed a multitude of advanced acquisition sequences, providing researchers with more resolution options. This progress, while beneficial, has consequently introduced a critical challenge for deep learning methods: data aggregated from various sources are now characterized by substantial variability in spatiotemporal resolution. Although all fMRI data is inherently 4D (3D spatial information across a time dimension), this resolution heterogeneity is pronounced.
Mainstream public datasets, for instance, exhibit spatial resolutions varying significantly from 2$mm$ to 4$mm$, with temporal sampling rates ranging drastically from 720$ms$ to over 3000$ms$~\cite{van2013wu,di2014autism,adhd2012adhd,Satterthwaite2016ThePN}. Moreover, anatomical differences between subjects mean the explicit 4D data shape (i.e., voxel count) varies across individuals, further compounding this personalized variability.

To achieve the unified input required by deep learning models, the current mainstream fMRI pipeline typically addresses inherent data heterogeneity through comprehensive preprocessing. Despite the development of numerous distinct libraries, the core preprocessing methodology remains fundamentally similar~\cite{esteban2019fmriprep,bellec2017neuro,craddock2013neuro,glasser2013minimal}. As shown in~\cref{fig:motivation}, the multi-stage preprocessing pipeline first converts the \textit{Raw fMRI} series into \textit{Native fMRI} through rapid operations such as slice-timing correction and head-motion correction, which are performed in each subject's original scanner space and typically require only a few minutes per subject. Second, much more computationally demanding stage, the pipeline incorporates the subject's structural MRI (sMRI) to handle inter-subject anatomical variability: the sMRI undergoes intensive anatomical processing and the functional data are then non-linearly registered to a common standard template (e.g., the MNI152 template), a procedure that often takes hours. This yields the \textit{Template fMRI}, in which all subjects' 4D fMRI volumes are warped into the same template-defined coordinate system and thus share an identical spatial structure, i.e., the same voxel grid, spatial layout, and standardized resolution as the template brain. Deep learning models are subsequently trained either directly on these normalized 4D volumes~\cite{wang2026omni,wang2025slim,Peng2025WholebrainTR,Sun2025VoxelLevelBS,kwon2025predicting,kim2023swift} or on dimensionality-reduced \textit{ROI-based fMRI} representations derived using brain parcellation atlases in template space~\cite{caro2023brainlm,dong2024brain,xia2026brain,dong2025brain,yang2024brainmass}.

While full preprocessing pipelines make deep learning on fMRI technically feasible, they introduce several structural limitations. First, model outputs are highly sensitive to preprocessing software and parameter choices, and different pipelines can yield systematically different representations. Second, mapping all subjects into a predefined template space inevitably suppresses subject-specific anatomical variability and introduces interpolation and partial-volume artifacts. Third, such pipelines are computationally intensive and difficult to scale. In the era of large foundation models trained on thousands or tens of thousands of subjects, repeated normalization and resampling become both economically and scientifically inefficient~\cite{caro2023brainlm,yang2024brainmass,dong2024brain}.
A natural alternative is to operate directly on native-space fMRI. However, existing approaches typically rely on subject-specific projection networks to map native voxels into a shared latent space~\cite{scotti2023reconstructing,scotti2024mindeye2}. While effective at small scale, this strategy prevents parameter sharing across heterogeneous datasets and becomes impractical for large-cohort foundation modeling.

In this paper, we introduce \textbf{FlexiBrain}, a resolution-adaptive voxel-level encoding framework for native fMRI. Instead of defining patches by fixed voxel counts, FlexiBrain parameterizes patch sizes in real-world physical units (millimeters and seconds). This design ensures that each token corresponds to a comparable anatomical and temporal receptive field across datasets with different spatial resolutions and fields of view.
To enable parameter sharing without destructive resampling, we develop a resolution-aware dynamic patch resizing that continuously resizes embedding kernels according to the input voxel spacing. As a result, resolution heterogeneity no longer requires dataset-specific adapters or retraining, allowing scalable native-space foundation modeling.
To further reduce the computational burden of voxel-level modeling, we instantiate FlexiBrain with a purpose-designed Mamba-JEPA backbone. The Mamba architecture replaces quadratic self-attention with linear-complexity state-space modeling, enabling scalable processing of high-resolution 4D fMRI volumes.
Meanwhile, the JEPA formulation shifts the objective from full signal reconstruction to latent predictive learning. By masking large background regions and avoiding heterogeneous voxel-wise reconstruction losses, this design improves computational efficiency while focusing representation learning on informative spatiotemporal dynamics.
Across five diverse downstream neuroscience tasks, FlexiBrain consistently outperforms recent state-of-the-art methods, achieving performance gains of up to 12 percentage points, while operating directly in native space without template normalization.

Our main contributions are summarized as follows:

\begin{itemize}
\item We identify resolution heterogeneity as a fundamental scalability bottleneck for voxel-level fMRI foundation models, and propose a native-space framework that enables parameter sharing without template normalization or subject-specific adapters.

\item We introduce FlexiBrain, a resolution-adaptive tokenization scheme that defines patches in physical units and dynamically resizes embedding kernels, allowing consistent receptive fields across heterogeneous spatial resolutions.

\item Extensive experiments on five downstream diagnosis tasks demonstrate that FlexiBrain outperforms template-based methods while operating directly on native 4D fMRI data.
\end{itemize}

\section{Related work}

\paragraph{fMRI Preprocessing}

Recent pipelines for fMRI preprocessing follow a two-track routine~\cite{esteban2019fmriprep}: anatomical processing (bias-field correction, skull stripping, tissue segmentation, surface reconstruction) and functional BOLD processing (reference volume creation, motion/susceptibility correction, slice-timing correction, coregistration to anatomy, and resampling).  A practical design choice is the output space. Processing in each subject's \textit{native space} is faster (fewer and cheaper resampling/warps) and preserves individual geometry and idiosyncratic anatomy, which benefits subject-specific modeling. In contrast, transforming to a \textit{standard template space} (e.g., MNI) is slower due to non-linear normalization but unifies data into a common coordinate frame, simplifying batching, voxel correspondence, and cross-subject aggregation. Recent voxel-level models adopt the standard space for ease of model design (fixed shapes, consistent spatial priors) and benchmarking~\cite{kim2023swift,wang2025dca,kwon2025predicting}. Yet forcing all subjects into a common template via non-linear warps inevitably introduces interpolation error and partial-volume artifacts, and can suppress subject-specific spatial variation (e.g., regional size/shape), potentially biasing downstream analyses.
Beyond voxel-level methods, ROI-based fMRI represents the most widely adopted format in contemporary deep learning for neuroscience~\cite{caro2023brainlm,yang2024brainmass,dong2024brain,dong2025brain,ye2023explainable}. This format reduces the fMRI volume to a 2D representation (channels $\times$ time), where each channel contains the average signal within a specific Region-of-Interest (ROI) defined by a brain parcellation atlas. Since these atlases are typically designed and validated in the standard template space (e.g., MNI), ROI-based fMRI must inherently be derived from Template fMRI. Consequently, this highly popular data format inherits all the aforementioned disadvantages associated with non-linear normalization and damages fine-grained spatial information.

\paragraph{Deep Learning in Native fMRI}

While Native fMRI analysis is favored in traditional neuroscience for its ability to minimize information loss and preserve subject-specific anatomy, its usage in large-scale deep learning remains challenging due to significant cross-site and inter-subject heterogeneity. Data exhibit stark differences in acquisition protocols, such as the spatial/temporal resolution gap between ABIDE ($\approx 3mm,2000ms$)~\cite{di2014autism} and UK Biobank ($\approx 2.4mm,735ms$)~\cite{bycroft2018uk}. Furthermore, hardware evolution promises an influx of increasingly finer-grained spatiotemporal data, exacerbating this resolution diversity. This heterogeneity has traditionally prevented the application of a single, unified deep learning model, forcing most works into the restrictive standard template space.
Research directly leveraging Native fMRI is scarce and largely confined to specialized tasks or small-sample cohorts. For example, recent fMRI-to-image reconstruction models include MindEye2~\cite{scotti2023reconstructing,scotti2024mindeye2}, which focuses on subject-specific mapping.

These methods select crucial voxels based on individual response features and train a dedicated mapping from the subject's native voxels to a shared latent space (e.g., CLIP embeddings) for each participant. While effective for individual decoding, this approach does not scale to variable resolution 4D fMRI from large, diverse cohorts.

\section{Method}

\begin{figure}[t]
  \centering
   \includegraphics[width=\linewidth]{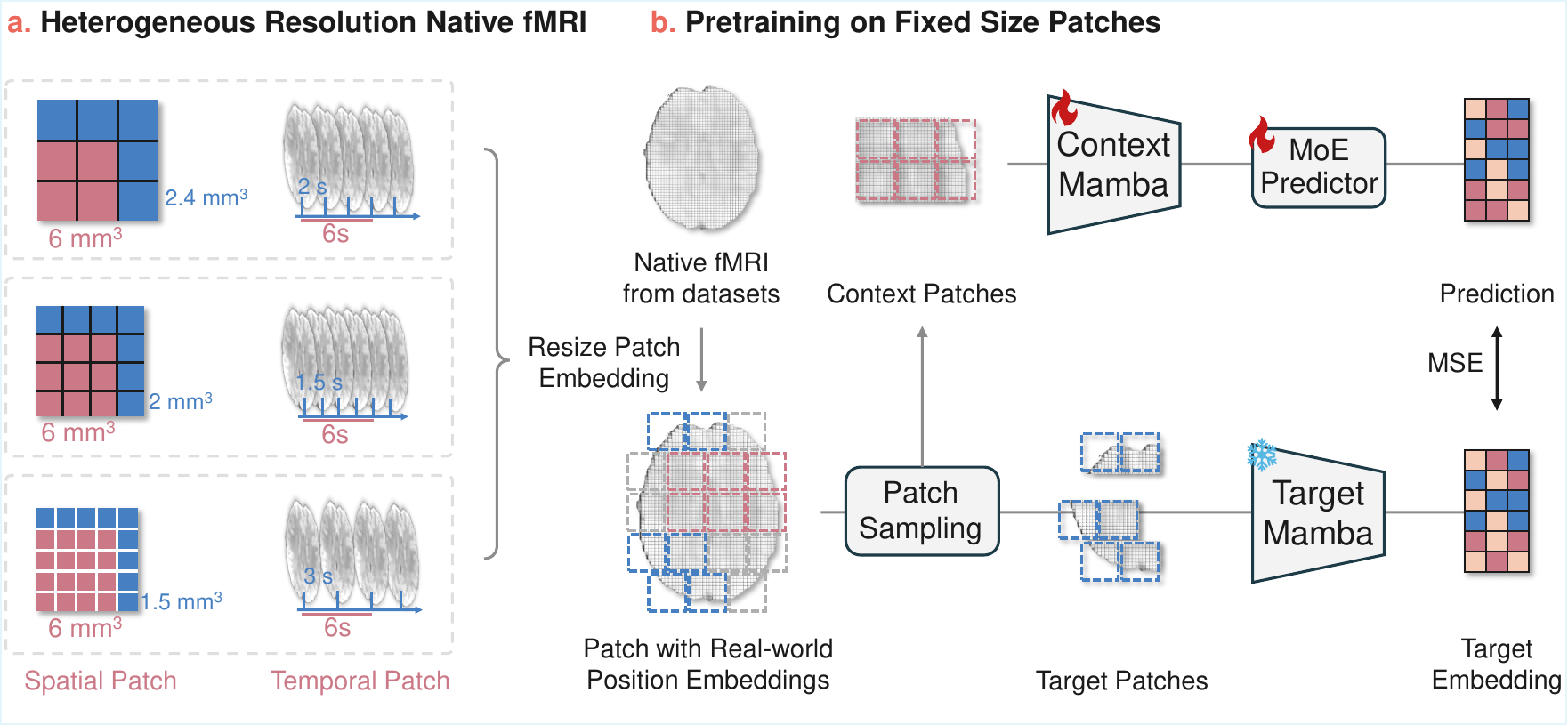}
   \caption{\textbf{FlexiBrain Pipeline.}
   \textbf{(a)} FlexiBrain accepts Native fMRI with  heterogeneous resolution from diverse datasets. To achieve resolution-agnosticism, the foundational spatial patch size (millimeters) and temporal patch size (seconds) are defined in real-world physical units. FlexiBrain uses a dynamic patch resizing to accommodate the variable number of voxels (due to differing resolutions) within a fixed physical patch size.
   \textbf{(b)} The physical patches with position embeddings in real-world coordinate system are sampled to create target and context patches for a self-supervised approach. The framework employs a novel Mamba-JEPA that is specifically designed to avoid background input and utilizes the Mixture-of-Experts module to mitigate representation collapse.
   }
   \label{fig:pipeline}
\end{figure}

Fig.~\ref{fig:pipeline} illustrates the overall framework of FlexiBrain. Heterogeneous fMRI data with varying spatial and temporal resolutions are tokenized using a dynamic patch resizing defined in physical units.
This mechanism ensures that patch tokens correspond to consistent physical receptive fields across heterogeneous fMRI acquisitions, enabling resolution-agnostic parameter sharing and stable representation learning directly in native space.
By resizing existing filters to the target physical scale rather than relearning resolution-specific ones, the tokenizer can be seamlessly transferred across datasets without reconfiguration.
The resulting fine-grained fMRI tokens are processed by a Mamba backbone, which enables efficient long-range modeling while significantly reducing computational cost. Training is guided by a JEPA-based proxy loss, which avoids heavy voxel-level reconstruction and mitigates loss imbalance arising from heterogeneous resolutions.

\subsection{Dynamic Patch Resizing}

\paragraph{Motivation.}
Voxel-level fMRI data exhibit substantial heterogeneity in spatial resolution and temporal sampling, making resolution-specific tokenizers brittle and difficult to transfer across datasets.
To address the heterogeneity in 4D fMRI data across varying spatial resolutions and temporal resolution (TR), we introduce a Dynamic Patch Resizing (Fig.~\ref{fig:pipeline} a).
Rather than relearning or discretely redefining filters for each resolution, we continuously resize a shared set of base kernels to the target physical scale, ensuring smooth parameter sharing across heterogeneous inputs. This continuous adaptation preserves the functional semantics of the tokenizer across resolutions, enabling direct transfer without reconfiguration or retraining.

\paragraph{Dynamic Patch Size Determination.}
Given a target physical temporal duration $\tau$ (e.g., 6.0 seconds) and target spatial resolution $\rho = (\rho_x, \rho_y, \rho_z)$ (e.g., 12 mm), the subject-specific patch sizes $k_t$ and $k_d$ as the discrete temporal and spatial patch dimensions are dynamically calculated based on the input's time resolution ($tr$) and voxel sizes $(v_x, v_y, v_z)$,$\left\lfloor  \right\rceil$ indicates nearest integer rounding:
\begin{equation}
k_t = \left\lfloor \frac{\tau}{tr} \right\rceil, \quad k_d = \left\lfloor \frac{\rho_d}{v_d} \right\rceil \quad \text{for } d \in \{x, y, z\}
\end{equation}

\paragraph{Continuous Kernel Resizing.}
Instead of randomly initializing distinct weights for every possible resolution, we maintain learnable base kernels $W_t^{(base)}$ for the temporal dimension and $W_{sp}^{(base)}$ for the spatial dimensions. To adapt these base kernels to the dynamically computed sizes $k_t$ and $(k_x, k_y, k_z)$, we utilize $\mathcal{R}^+$, a geometric operator that optimally projects the weights via the Moore-Penrose pseudo-inverse of the forward interpolation matrix to preserve their functional mapping~\cite{Beyer2022FlexiViTOM}:
\begin{align}
    W_t &= \mathcal{R}^+_{1D}\left(W_t^{(base)} \rightarrow k_t\right) \\
    W_{sp} &= \mathcal{R}^+_{3D}\left(W_{sp}^{(base)} \rightarrow (k_x, k_y, k_z)\right)
\end{align}

\paragraph{Time-to-Space Forward Tokenization.}
First, a 1D temporal convolution with weight $W_t$ and stride $k_t$ aggregates short-range temporal dynamics within each physical window, transforming the input
$X \in \mathbb{R}^{D \times H \times W \times T}$
into an intermediate feature map
$X_t \in \mathbb{R}^{D \times H \times W \times D_m \times T'}$,
where $T' = T / k_t$ and $D_m$ denotes the intermediate channel dimension.

To enable efficient spatial tokenization, we fold the reduced temporal dimension $T'$ into the channel dimension,
\[
\tilde{X}_t = \text{Reshape}(X_t) \in \mathbb{R}^{(D_m \cdot T') \times D \times H \times W},
\]
such that temporal information is encoded as channel-wise features.
Correspondingly, the spatial kernel $W_{sp}$ is repeated $T'$ times along its input channel dimension to match $\tilde{X}_t$, reflecting the assumption that spatial structure is invariant to temporal indexing. A 3D convolution with stride $(k_x, k_y, k_z)$ is then applied to project the spatial grid into a $D_{out}$-dimensional latent space, yielding the final joint spatiotemporal patch embeddings.
Finally, purely background patches are discarded using a subject-specific foreground mask, and the remaining valid tokens are flattened into a 1D sequence.

\paragraph{Continuous Physical 3D Positional Encoding.}
Discarding background patches disrupts the regular 3D grid, rendering standard discrete grid-based positional encodings ineffective, especially across subjects with varying spatial resolutions. To preserve precise anatomical awareness, we introduce a continuous physical 3D positional encoding.
First, we determine the exact physical center of each valid patch in the anatomical world space. Let $p_i \in \mathbb{R}^3$ denote the discrete voxel-space center index of the $i$-th patch. Using the subject-specific affine transformation matrix $A \in \mathbb{R}^{3 \times 3}$ and translation vector $t \in \mathbb{R}^3$ extracted from the NIfTI headers, the continuous physical coordinate $c_i \in \mathbb{R}^3$ is computed as:
$$
c_i = p_i A^\top + t
$$

To map these continuous physical coordinates into a high-dimensional representation, we apply a spatial scaling factor to mitigate unit variance and generate sinusoidal embeddings using logarithmically spaced frequency bands for each spatial axis. The concatenated features from all three axes are finally mapped to the target token dimension $D_{out}$ via a learnable linear projection. Finally, these positional embeddings are added to the flattened 1D spatial token sequence.

\subsection{Model Architecture: Mamba-JEPA}

\paragraph{Motivation.}
Our architectural design is motivated by the need to efficiently model ultra-long, fine-grained token sequences while learning robust semantic representations from noisy 4D fMRI data. Native-space voxel-level tokenization yields thousands of tokens per subject, rendering standard self-attention prohibitively expensive. Moreover, voxel-level reconstruction objectives are ill-suited for fMRI, as they emphasize high-frequency noise rather than meaningful functional structure. To address these challenges, we combine a linear-time state-space backbone (Mamba) with a latent predictive learning objective (JEPA), further augmented by structural asymmetry via a lightweight Mixture-of-Experts (MoE) to prevent representation collapse (Fig.~\ref{fig:pipeline} b).

\paragraph{Mamba for Ultra-Long Sequences.}
Let $\mathbf{X}\!\in\!\mathbb{R}^{D\times H\times W\times T}$ denote the input 4D fMRI signal. A dynamic patch resizing first tokenizes $\mathbf{X}$ into a sequence $\mathbf{Z}$ with real-world physical positional embeddings.

Even after aggressively discarding approximately 80\% of the purely background patches, the resulting token sequence $\mathbf{Z}$ still contains around 4,000 tokens per subject. Processing sequences of this magnitude with standard self-attention incurs an intractable $\mathcal{O}(N^2)$ computational overhead. To accelerate training while preserving the ability to model complex spatial-temporal dependencies, we replace the standard self-attention mechanism with the Selective State Space Model (Mamba), which operates in linear time $\mathcal{O}(N)$.

During pre-training, a random masking strategy partitions the token sequence into a visible set $\mathcal{V}$ and a masked set $\mathcal{M}$. The context encoder $f_{\theta}$, built upon stacked Mamba blocks, processes only the visible tokens $\mathbf{Z}_{\mathcal{V}}$.

\paragraph{Latent Predictive Objective.}
For the self-supervised proxy task, Masked Autoencoders (MAE) attempt to reconstruct masked tokens in the raw voxel space. However, fMRI data is notoriously noisy; forcing the network to reconstruct high-frequency voxel-level noise wastes computational capacity and degrades semantic representation quality. This issue is amplified in native-space, resolution-heterogeneous fMRI, where a reconstruction decoder would need to balance losses across different voxel grids and acquisition protocols. Meanwhile, given the variable resolution of fMRI inputs, fine-grained data would disproportionately dominate the MAE loss without appropriate weighting. Therefore, we adopt JEPA, which shifts the predictive reconstruction strictly into the semantic latent space.

The target encoder $f_{\xi}$ (sharing the same structural design as $f_{\theta}$) processes the ground-truth masked tokens $\mathbf{Z}_{\mathcal{M}}$ to generate latent targets $\mathbf{H}_{\mathcal{M}}$. To stabilize learning, the target encoder's parameters $\xi$ are updated via an exponential moving average (EMA) of the context encoder's parameters $\theta$ and $\mu$ is the weight:
$$
\xi \leftarrow \mu\xi + (1-\mu)\theta
$$

\paragraph{Asymmetric Encoding via MoE to Prevent Collapse.}
A fundamental vulnerability of JEPA is representation collapse, where the encoder trivializes the objective by mapping all inputs to a constant vector. To explicitly combat this, we introduce structural asymmetry between the encoding pathways by appending a lightweight MoE module $h_{\psi}$ to the tail of the context encoder. The MoE is not intended to replace JEPA's standard asymmetric context--target pathways or the EMA target encoder; rather, it complements them by adding token-adaptive transformations on the context side. This architectural asymmetry prevents trivial solutions by ensuring that the predictive pathway cannot collapse to a constant mapping shared by both encoders. Instead of relying on complex routing equations, the MoE module simply employs a token-level router to adaptively compute soft mixture weights, routing each visible token to a subset of MLP-based expert networks. This yields enriched, token-adaptive context representations $\widehat{\mathbf{H}}_{\mathcal{V}} = h_{\psi}(f_{\theta}(\mathbf{Z}_{\mathcal{V}}))$.

\paragraph{State-Space Prediction and Loss.}
To perform the predictive task, we construct a full sequence by scattering the MoE-processed visible tokens $\widehat{\mathbf{H}}_{\mathcal{V}}$ back to their original sequence positions and inserting a learned mask token $\mathbf{m}$ at the masked locations $\mathcal{M}$. A shallow, Mamba-based predictor $g_{\phi}$ then diffuses information over this sequence to yield latent predictions $\widehat{\mathbf{H}}_{\mathcal{M}}$ at the masked locations.

We train the model by minimizing the $\ell_2$ distance between the normalized predictions and the EMA-generated targets:
$$
\mathcal{L} = \frac{1}{|\mathcal{M}|} \sum_{i \in \mathcal{M}} \left\| \frac{\widehat{\mathbf{H}}_i}{\|\widehat{\mathbf{H}}_i\|_2} - \frac{\mathbf{H}_i}{\|\mathbf{H}_i\|_2} \right\|_2^2
$$

\paragraph{Downstream Adaptation.}
For downstream inference, the pre-trained Mamba backbone extracts feature representations from the input tokens. A learnable class token is prepended to this sequence, and the concatenated representation is processed by a lightweight transformer classifier head. Finally, the feature corresponding to the class token is normalized and passed through an MLP to yield the final prediction logits.

\section{Experiments}

\subsection{Data Preprocessing}

We leveraged self-supervised training on three public neuroimaging datasets: the Autism Brain Imaging Data Exchange (ABIDE) \cite{di2014autism}; the Alzheimer's Disease Neuroimaging Initiative (ADNI)~\cite{jack2008alzheimer}; the ADHD-200 Sample (ADHD) \cite{adhd2012adhd} and used the Parkinson progression marker initiative (PPMI) ~\cite{marek2011parkinson} as an external dataset to assess the model's generalization performance.
These are all \textbf{multi-site datasets} comprising images collected across diverse centers and protocols; consequently, the temporal and spatial resolutions vary significantly across the cohorts. Details of datasets (e.g., voxel spacing and TR distribution) and task can be found in \textbf{Appendix C}.
This exclusion was applied solely to match the subject set used by baselines that require structural scans, ensuring a fair comparison.
For FlexiBrain,

all images retain their temporal and spatial resolution. We further verified that neither temporal resolution nor spatial resolution was associated with the class labels, indicating that resolution heterogeneity does not induce shortcut learning or spurious label--resolution confounds.
All images for baselines were preprocessed following the specific requirements in the corresponding  paper.

\begin{table*}[t]
\centering

\renewcommand{\arraystretch}{1.3}
\caption{Mean accuracy and standard deviations of predictions on disorder diagnosis over three random seeds. The best results are highlighted in \textbf{bold}, and second-best results are \underline{underlined}.  MCI: Mild cognitive impairment; AD: Alzheimer's disease. * denotes a large effect size with Cohen's d $\ge$ 0.8.}
\label{tab:ndd-comparison}

\newcommand{\std}[1]{$_{\pm#1}$}

\resizebox{\textwidth}{!}{
\begin{tabular}{l*{10}{c}}
\toprule
\multirow{2}{*}{\textbf{Model}} &

\multicolumn{2}{c}{\textbf{ADHD-200}} & \multicolumn{2}{c}{\textbf{ABIDE}} & \multicolumn{2}{c}{\textbf{ADNI (MCI)}} & \multicolumn{2}{c}{\textbf{ADNI (AD)}} & \multicolumn{2}{c}{\textbf{PPMI}} \\
\cmidrule(lr){2-3}\cmidrule(lr){4-5}\cmidrule(lr){6-7}\cmidrule(lr){8-9}\cmidrule(lr){10-11}
 & ACC\% & F1\% & ACC\% & F1\% & ACC\% & F1\% & ACC\% & F1\% & ACC\% & F1\% \\
\midrule

BrainNetCNN~\cite{kawahara2017brainnetcnn}  & 54.46\std{2.47} & 53.07\std{3.89} & 57.90\std{3.25} & 57.84\std{3.21} & 59.74\std{4.50} & 56.48\std{6.04} & 74.01\std{3.91} & 66.47\std{4.18} & 64.24\std{2.41} & 58.15\std{1.44} \\
BrainGNN~\cite{li2021braingnn}  & 55.87\std{3.88} & 54.93\std{4.46} & 54.67\std{3.72} & 53.45\std{3.28} & 63.20\std{7.38} & 58.36\std{12.50} & 78.53\std{2.59} & 72.03\std{5.38} & 55.56\std{4.81} & 51.31\std{2.99} \\
BrainMass~\cite{yang2024brainmass}  & 60.78\std{0.49} & \second{60.05\std{0.79}} & \second{63.33\std{1.18}} & \best{65.35\std{2.00}} & 62.39\std{0.60} & 61.35\std{0.60} & 73.45\std{3.91} & 62.10\std{0.70} & 58.68\std{4.21} & 51.95\std{2.23} \\
Brain-JEPA~\cite{dong2024brain}  & 59.74\std{0.23} & 55.20\std{6.40} & 59.74\std{2.64} & 54.94\std{11.98} & 64.53\std{0.60} & 55.17\std{0.66} & 80.23\std{0.80} & \second{79.66\std{1.03}} & 47.91\std{3.61} & 48.60\std{2.42} \\
BrainHarmonix-F~\cite{dong2025brain} & 59.74\std{1.43} & 58.76\std{1.52} & 54.03\std{0.55} & 43.04\std{7.12} & 60.61\std{0.75} & 51.86\std{5.89} & 76.84\std{0.98} & 71.17\std{1.55} & 65.63\std{3.61} & 55.19\std{2.42} \\

SwiFT~\cite{kim2023swift}   & \second{62.50\std{1.50}} & 59.99\std{4.71} & 61.02\std{0.28} & 61.36\std{3.05} & 65.38\std{1.29} & \second{72.54\std{0.16}} & 74.90\std{1.88} & 80.76\std{0.33} & 61.55\std{2.52} & 58.23\std{4.20} \\
NeuroSTORM~\cite{li2025towards} & 59.51\std{1.01} & 59.05\std{1.01} & 54.52\std{2.40} & 53.04\std{0.70} & \second{66.67\std{1.06}} & 62.17\std{8.93} & \best{84.26\std{0.80}} & \best{83.91*\std{0.75}} & \second{68.25\std{1.18}} & \second{65.18\std{1.19}}  \\
\rowcolor{rowgray}
FlexiBrain & \best{69.76*\std{1.75}}\label{special} & \best{69.85*\std{1.69}} & \best{65.45*\std{0.69}} & \second{64.82\std{0.52}} & \best{79.16*\std{1.48}} & \best{78.82*\std{1.51}} & \second{82.34\std{2.39}} & 79.41\std{4.42} & \best{74.31*\std{1.77}} & \best{70.54*\std{1.19}} \\
\bottomrule
\end{tabular}
}
\end{table*}

\subsection{Disease Diagnosis}

To validate the robustness of our model to heterogeneous data resolutions, we curated a mixed-resolution dataset for self-supervised pre-training by combining three public fMRI sources: ABIDE, ADHD-200, and ADNI. For downstream evaluation, the pretrained model was fine-tuned and benchmarked against multiple baselines on four tasks, using accuracy (ACC) and F1-score as evaluation metrics. Specifically, ABIDE and ADHD-200 were formulated as binary classification tasks, while ADNI was evaluated using two clinically meaningful binary settings: (i) mild cognitive impairment (MCI) vs. cognitively normal (CN), and (ii) Alzheimer's disease (AD) vs. CN.
To explicitly assess out-of-distribution generalization, we further evaluated the pretrained model on the external, unseen PPMI dataset, where a three-class classification task was performed. Even compared to large-scale pretraining baselines~\cite{yang2024brainmass,dong2024brain,dong2025brain,li2025towards}, our model consistently achieved strong performance as reported in Table~\ref{tab:ndd-comparison}. Detailed hyperparameter settings, and implementation details for both our model and the baselines are provided in \textbf{Appendices A and B}.
We further audited this result under acquisition heterogeneity. PPMI is treated as a leave-one-dataset-out evaluation because it is not used during pretraining; when stratified by voxel spacing, FlexiBrain obtains 74.40\% weighted F1 on the dominant higher-resolution bin and 66.96\% weighted F1 on the lower-resolution bin (Supplementary Table~\ref{tab:supp_ppmi_resolution}). A controlled PPMI downsampling study shows that performance declines under spatial and temporal degradation, especially when both are degraded, indicating that the physical-unit tokenizer shares parameters across resolutions without collapsing all resolutions into an equivalent representation (Supplementary Table~\ref{tab:supp_downsampling}). We also include gender prediction on the heterogeneous ABIDE and ADHD-200 cohorts as an additional downstream check (Supplementary Table~\ref{tab:supp_gender}). A stricter ABIDE leave-one-scanner-center-out analysis degrades when the held-out scanner center is unseen during both pretraining and downstream training; thus, FlexiBrain addresses resolution heterogeneity, but scanner-center domain shift remains an open challenge.

\subsection{Choice of Preprocessing}

\begin{figure}[t]
  \centering
  \includegraphics[width=\linewidth]{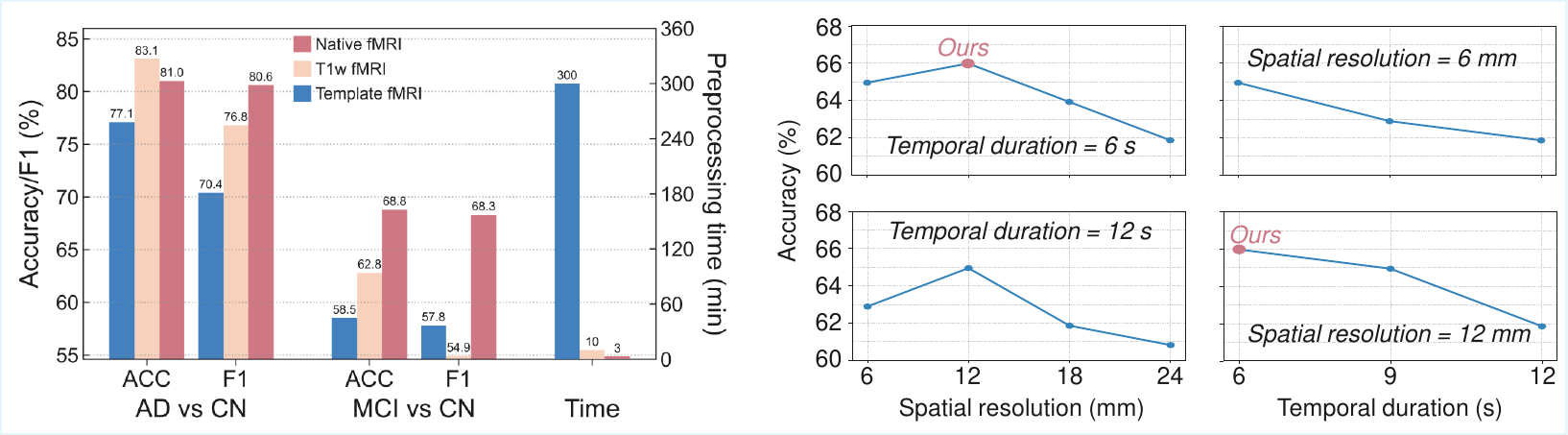}
  \caption{\textbf{Left: Ablation of preprocessing steps} on ADNI dataset. Comparison of classification accuracy and preprocessing time between three fMRI states ( Native fMRI, T1w fMRI, Template fMRI), with models trained from scratch. \textbf{Right: Ablation study on patch size.} The figure shows the accuracy of classification on ADHD dataset under different temporal duration and spatial resolution. Our settings with 6 s and 12 mm show best performance. AD: Alzheimer's disease; MCI: Mild cognitive impairment; CN: Cognitively normal.}
  \label{fig:2+3}
\end{figure}

\paragraph{The effect of the transformation on the predictions.}
Here we used different preprocessing states of fMRI to compare their performance by training the models from scratch. (i) Native fMRI / minimal \textbf{(ours)}: brain extraction only; head-motion parameters and slice-timing information are estimated, and the series remains in native space. (ii) T1w (weighted) fMRI: the native fMRI is first affinely registered to each subject's T1w anatomical scan, and not nonlinearly warped into a population template.  (iii) Template fMRI (standard space): a nonlinear warp maps the data to a standard template, which is the most common preprocessing choice to align subjects into a standard space under heterogeneous resolutions.
As shown in Fig.~\ref{fig:2+3} Left, the native-space approach achieves the best performance while dramatically reducing preprocessing time from around 5 hours to 3 minutes. We hypothesize that the inferior performance of template fMRI is because spatial normalization reduces inter-subject variability and information entropy, while the requisite interpolation introduces noise artifacts.

\paragraph{Ablation of Patch Size.}
We systematically ablate the temporal duration \(\tau\) and the spatial resolution \(\rho\) (Fig.~\ref{fig:2+3} Right). We evaluated performance with ADHD by training it from scratch and find the best settings with $\tau = 6$ seconds and $\rho = 12$ millimeters.  Increasing \(\tau\) beyond 6 seconds and \(\rho\) beyond 12 millimeters leads to coarser patches that blur details, while very small \(\tau\) or voxel spacings cause discretization, reducing sensitivity to subtle differences and harming generalization across protocols.

\subsection{Choice of Architecture}

\paragraph{Effectiveness of Mamba over ViT.}
To validate our backbone choice, we compare the Mamba backbone against a standard attention-based ViT under a comparable parameter budget ($\sim$50M). Profiling results reveal that the ViT backbone incurs substantial computational and memory overheads (\textbf{28.24} GFLOPs, \textbf{8.29} GB memory). In contrast, Mamba significantly reduces the resource footprint to just \textbf{18.59} GFLOPs and \textbf{1.97} GB memory. Beyond these critical efficiency gains, we further evaluated both pre-trained architectures on downstream tasks. As shown in Table~\ref{tab:ablation}, the Mamba backbone achieves superior downstream performance. We attribute this empirical advantage to Mamba's inherent capacity for modeling ultra-long, irregular voxel sequences (nearly 4,000 tokens), effectively bypassing the quadratic context bottleneck that degrades representation learning in standard self-attention mechanisms.

\begin{table*}[t]
\centering

\renewcommand{\arraystretch}{1.18}
\caption{Ablation study on architecture. We compared four downstream tasks under different settings. The best results are highlighted in bold. MCI: Mild cognitive impairment; AD: Alzheimer's disease.}
\label{tab:ablation}
\resizebox{\textwidth}{!}{
\begin{tabular}{ccccccccccccc}
\toprule
 \multirow{2}{*}{\textbf{Backbone}} & \multirow{2}{*}{\textbf{Pre-training}} & \multirow{2}{*}{\textbf{MoE}} &
\multicolumn{2}{c}{\textbf{ABIDE}} & \multicolumn{2}{c}{\textbf{ADHD-200}} & \multicolumn{2}{c}{\textbf{ADNI (MCI) }} & \multicolumn{2}{c}{\textbf{ADNI (AD)}} \\
\cmidrule(lr){4-5}\cmidrule(lr){6-7}\cmidrule(lr){8-9}\cmidrule(lr){10-11}
  &  &   & ACC\% & F1\% & ACC\% & F1\% & ACC\% & F1\% & ACC\% & F1\% \\
\midrule
 Mamba & \cmark & \xmark & 52.55 & 52.32 & 61.86 & 58.76 & 71.88 & 71.63 & 71.43 & 70.40 \\
 Mamba & \xmark & \cmark & 59.85 & 59.66 & 65.98 & 61.37 & 68.75 & 68.25 & 80.95 & \best{80.59} \\
 ViT & \cmark & \cmark & 61.31 & 60.41 & 63.92 & 59.03 & 66.66 & 66.60 & 76.85 & 72.92 \\
\rowcolor{rowgray}
 Mamba & \cmark & \cmark & \best{65.45} & \best{64.82} & \best{69.76} & \best{69.85} & \best{79.16} & \best{78.82} & \best{82.34} & 79.41 \\
\bottomrule
\end{tabular}
}
\end{table*}

\begin{wrapfigure}[17]{r}{0.5\textwidth}
  \centering
   \includegraphics[width=\linewidth]{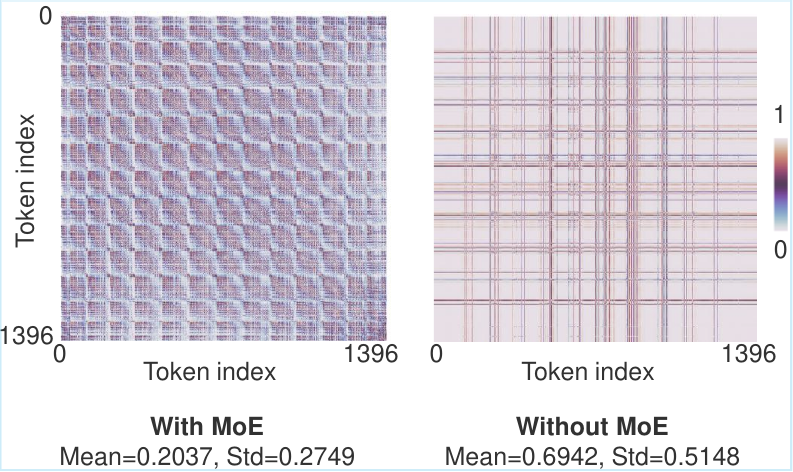}
   \caption{
\textbf{Target embedding similarity matrix}  exhibits strong vertical/horizontal streaks, indicating low-rank, directional convergence to a few dominant directions.}
   \label{fig:similarity}
\end{wrapfigure}

\paragraph{MoE and Pre-training.}
We compared the performance of our model with and without the MoE and pre-training. As shown in Table~\ref{tab:ablation}, there is a noticeable decline in model performance when the MoE module or the pre-training stage is removed.
Pre-training on diverse data allows the model to learn a robust and generalizable feature representation. This representation serves as an initial distribution for fine-tuning, which in turn significantly boosts performance and data efficiency on downstream tasks.
MoE effectively prevents model collapse. We visualized the tokens output by the target encoder (ordered by the position of tokens in the flattened 1D sequence of valid patch tokens) in cosine-similarity map (Fig.~\ref{fig:similarity}), which reveals a marked difference in representational geometry. With MoE, the mean similarity drops substantially and the pattern becomes fine-grained and predominantly near-diagonal, indicating content-dependent structure rather than convergence to a few dominant directions. Additional experiments are provided in \textbf{Appendix B}.

\begin{wrapfigure}[22]{r}{0.48\textwidth}
  \centering
   \includegraphics[width=\linewidth]{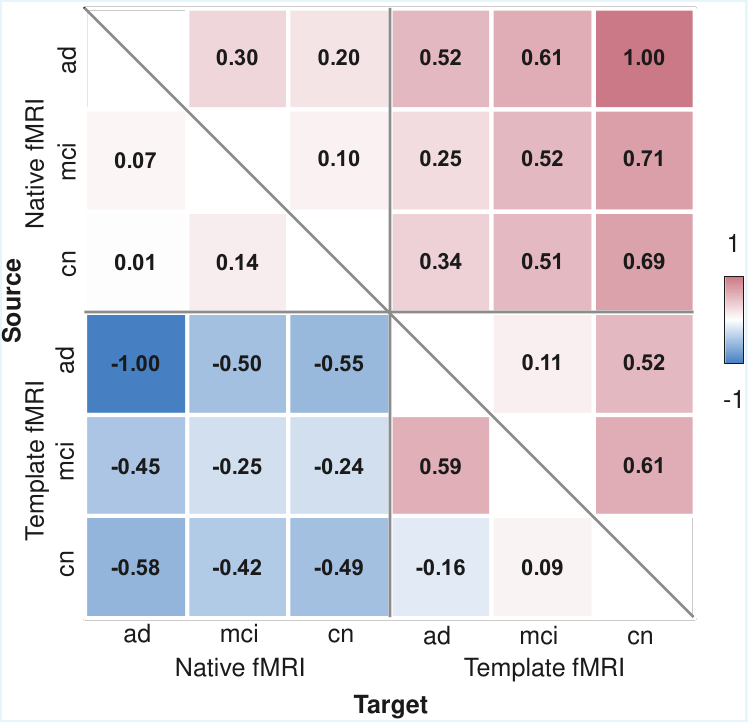}
   \caption{
\textbf{Transfer task} illustrates the normalized domain relations and disease-dependent transferability. Warm color denotes higher transferability, while cool colors indicate weaker transfer.}
   \label{fig:4+5}
\end{wrapfigure}

\subsection{Interpretability}

Transfer learning improves performance on a target task by leveraging knowledge acquired from a related source task. In general, stronger transfer gains indicate a closer relationship between the source and target tasks, thereby revealing their underlying structural and semantic connections~\cite{qu2024uncovering,zamir2018taskonomy,qu2022transfer}. Following this principle, we employ a taskonomy-based analysis to investigate whether native-space fMRI data capture richer individual-specific and structural information than template-space data, leading to improved transferability.

Specifically, we train six source models using ADNI data across two spatial domains---Native space (FSL) and Template space (MNI)---and three diagnostic groups (AD, MCI, CN). This results in 30 directed source-to-target transfer experiments. All models share the same architecture and training objective as used during pretraining. For each ordered task pair $(s, t)$, we freeze the encoder pretrained on the source task and fine-tune a lightweight decoder on a small subset of the target task for 30 epochs, yielding a transfer loss $L_{s\to t}$, where lower values indicate stronger transferability. For visualization, we apply global min--max normalization to map transfer scores to $[-1, 1]$ and render the resulting taskonomy matrix as a heatmap (Fig.~\ref{fig:4+5}).

The taskonomy reveals clear relationships between spatial domain choice and diagnostic transferability. Cross-domain transfers from Native-space fMRI (FSL) to Template-space fMRI (MNI) outperform the reverse direction, suggesting that native-space representations preserve richer individual-specific and structural information, whereas template normalization improves spatial alignment at the cost of smoothing subject-specific signals. Within-domain transfers, both in Native and Template spaces, exhibit consistently lower transfer losses, indicating strong internal consistency under a shared spatial processing framework.

Across both within- and cross-domain transfers, AD demonstrates the strongest transferability, likely due to its pronounced and diverse pathological patterns. MCI shows intermediate transfer performance, consistent with its transitional clinical nature, while CN exhibits weaker cross-diagnostic transferability, reflecting more stable brain activity and reduced signal variability. Together, these results indicate that native-space representations contain richer transferable information and that disease-related heterogeneity plays a key role in determining transferability across diagnostic tasks.

\section{Discussion}

The transition from ROI-based analysis to fine-grained voxel-level modeling represents a natural evolution in fMRI research, analogous to the shift from static images to video modeling in computer vision. However, current voxel-level pipelines remain tightly coupled to heavy preprocessing, particularly spatial normalization to a common template. While such alignment enables cross-subject comparison, it also introduces structural and scalability constraints.

First, template normalization inevitably alters subject-specific anatomical variability through interpolation and smoothing, potentially degrading analyses that depend on fine-grained structural fidelity. Second, the computational cost and diversity of preprocessing standards limit reproducibility and scalability. These issues become critical in the era of foundation models, where training on tens of thousands of subjects makes repeated large-scale normalization economically and practically prohibitive. As open neuroimaging datasets continue to grow, this scalability bottleneck becomes increasingly prominent.

In this work, we propose FlexiBrain, a native-space voxel-level architecture that shifts the burden of alignment from data preprocessing to model design. The central challenge of native fMRI lies in resolution heterogeneity across datasets. While temporal inconsistencies have been partially addressed in prior work~\cite{dong2025brain}, voxel-level spatial heterogeneity has remained largely unexplored.

FlexiBrain addresses this challenge at the tokenization level. By defining patches in physical units (millimeters and seconds) and dynamically resizing embedding kernels, the model preserves consistent anatomical and temporal receptive fields across heterogeneous resolutions. Real-world coordinate positional encoding further enforces semantic consistency in native space, enabling parameter sharing without destructive resampling.

To make voxel-level modeling computationally feasible, we instantiate FlexiBrain with a Mamba-JEPA backbone. Linear-complexity state-space modeling replaces quadratic self-attention, enabling scalable processing of high-resolution 4D volumes. Meanwhile, the JEPA formulation shifts learning from voxel-wise reconstruction toward latent predictive objectives, avoiding heavy decoders and focusing representation learning on informative spatiotemporal dynamics. Across five downstream neuroscience tasks, FlexiBrain consistently outperforms recent state-of-the-art methods while operating directly in native space.

\paragraph{Limitations and Future Work.}

Although FlexiBrain demonstrates the feasibility of template-free native-space modeling, several limitations remain.
First, our experiments focus on disease-specific downstream tasks rather than training a large-scale foundation model. While the resolution-adaptive design is inherently suited to heterogeneous cohorts, validating its effectiveness in a truly large-scale pretraining setting remains future work.
Second, we operate on minimally preprocessed native fMRI (e.g., motion-corrected) to ensure temporal consistency within sessions. Extending the framework to raw acquisition data would move closer to a fully end-to-end learning paradigm, potentially reducing reliance on traditional preprocessing pipelines and enabling the utilization of data currently excluded during quality control.
Finally, while FlexiBrain enables parameter sharing across resolutions, acquisition-specific artifacts and scanner variability remain open challenges for native-space modeling. Our leave-one-scanner-center-out analysis indicates that resolution-adaptive tokenization alone does not eliminate domain shift when a scanner center is absent from both pretraining and downstream training. Addressing these factors may further enhance robustness and cross-dataset generalization. Future work should combine physical-unit tokenization with explicit site harmonization, nuisance auditing, or domain-adaptive pretraining for deployment across unseen acquisition centers.

\section{Conclusion}

We present FlexiBrain, a resolution-adaptive voxel-level framework for native fMRI modeling. By redefining tokenization in physical units and introducing dynamic patch resizing, FlexiBrain removes resolution heterogeneity as a barrier to parameter sharing across datasets. Combined with an efficient Mamba-JEPA backbone, the framework enables scalable representation learning directly in native space without template normalization.

Our results demonstrate that native-space modeling can achieve superior performance while reducing computational overhead, suggesting a promising direction for future fMRI foundation models. We hope this work encourages a shift from data-centric alignment toward model-centric adaptation in large-scale neuroimaging research.

\bibliographystyle{splncs04}
\bibliography{main}

@String(CVPR  = {IEEE Conf. Comput. Vis. Pattern Recog.})

@String(CVPR  = {CVPR})

@article{li2025towards,
  title={Towards a general-purpose foundation model for fMRI analysis},
  author={Li, Xiang and Wang, Cheng and Jiang, Yu and PENG, Zhihao and Li, Chenxin and Bang, Changbae and Zhao, Lin and Lv, Jinglei and Sepulcre, Jorge and Yang, Carl and others},
  year={2025}
}

@article{wang2026omni,
  title={Omni-fMRI: A Universal Atlas-Free fMRI Foundation Model},
  author={Wang, Mo and Ye, Wenhao and Xia, Junfeng and Zhang, Junxiang and Pan, Xuanye and Xu, Minghao and Deng, Haotian and Wen, Hongkai and Liu, Quanying},
  journal={arXiv preprint arXiv:2601.23090},
  year={2026}
}

@article{wang2025slim,
  title={SLIM-Brain: A Data-and Training-Efficient Foundation Model for fMRI Data Analysis},
  author={Wang, Mo and Xia, Junfeng and Ye, Wenhao and Liu, Enyu and Peng, Kaining and Feng, Jianfeng and Liu, Quanying and Wen, Hongkai},
  journal={arXiv preprint arXiv:2512.21881},
  year={2025}
}

@article{xia2026brain,
  title={Brain-DiT: A Universal Multi-state fMRI Foundation Model with Metadata-Conditioned Pretraining},
  author={Xia, Junfeng and Ye, Wenhao and Pan, Xuanye and Shen, Xinke and Wang, Mo and Liu, Quanying},
  journal={arXiv preprint arXiv:2604.12683},
  year={2026}
}

@article{bycroft2018uk,
  title={The UK Biobank resource with deep phenotyping and genomic data},
  author={Bycroft, Clare and Freeman, Colin and Petkova, Desislava and Band, Gavin and Elliott, Lloyd T and Sharp, Kevin and Motyer, Allan and Vukcevic, Damjan and Delaneau, Olivier and O’Connell, Jared and others},
  journal={Nature},
  volume={562},
  number={7726},
  pages={203--209},
  year={2018},
  publisher={Nature Publishing Group UK London}
}

@article{caro2023brainlm,
  title={BrainLM: A foundation model for brain activity recordings},
  author={Caro, Josue Ortega and Fonseca, Antonio H de O and Averill, Christopher and Rizvi, Syed A and Rosati, Matteo and Cross, James L and Mittal, Prateek and Zappala, Emanuele and Levine, Daniel and Dhodapkar, Rahul M and others},
  journal={bioRxiv},
  pages={2023--09},
  year={2023},
  publisher={Cold Spring Harbor Laboratory}
}

@article{di2014autism,
  title={The autism brain imaging data exchange: towards a large-scale evaluation of the intrinsic brain architecture in autism},
  author={Di Martino, Adriana and Yan, Chao-Gan and Li, Qingyang and Denio, Erin and Castellanos, Francisco X and Alaerts, Kaat and Anderson, Jeffrey S and Assaf, Michal and Bookheimer, Susan Y and Dapretto, Mirella and others},
  journal={Molecular psychiatry},
  volume={19},
  number={6},
  pages={659--667},
  year={2014},
  publisher={Nature Publishing Group}
}

@article{dong2024brain,
  title={Brain-jepa: Brain dynamics foundation model with gradient positioning and spatiotemporal masking},
  author={Dong, Zijian and Li, Ruilin and Wu, Yilei and Nguyen, Thuan Tinh and Chong, Joanna and Ji, Fang and Tong, Nathanael and Chen, Christopher and Zhou, Juan Helen},
  journal={Advances in Neural Information Processing Systems},
  volume={37},
  pages={86048--86073},
  year={2024}
}

@article{kim2023swift,
  title={Swift: Swin 4d fmri transformer},
  author={Kim, Peter and Kwon, Junbeom and Joo, Sunghwan and Bae, Sangyoon and Lee, Donggyu and Jung, Yoonho and Yoo, Shinjae and Cha, Jiook and Moon, Taesup},
  journal={Advances in Neural Information Processing Systems},
  volume={36},
  pages={42015--42037},
  year={2023}
}

@article{scotti2024mindeye2,
  title={Mindeye2: Shared-subject models enable fmri-to-image with 1 hour of data},
  author={Scotti, Paul S and Tripathy, Mihir and Villanueva, Cesar Kadir Torrico and Kneeland, Reese and Chen, Tong and Narang, Ashutosh and Santhirasegaran, Charan and Xu, Jonathan and Naselaris, Thomas and Norman, Kenneth A and others},
  journal={arXiv preprint arXiv:2403.11207},
  year={2024}
}

@article{scotti2023reconstructing,
  title={Reconstructing the mind's eye: fmri-to-image with contrastive learning and diffusion priors},
  author={Scotti, Paul and Banerjee, Atmadeep and Goode, Jimmie and Shabalin, Stepan and Nguyen, Alex and Dempster, Aidan and Verlinde, Nathalie and Yundler, Elad and Weisberg, David and Norman, Kenneth and others},
  journal={Advances in Neural Information Processing Systems},
  volume={36},
  pages={24705--24728},
  year={2023}
}

@article{kwon2025predicting,
  title={Predicting task-related brain activity from resting-state brain dynamics with fMRI Transformer},
  author={Kwon, Junbeom and Seo, Jungwoo and Wang, Heehwan and Moon, Taesup and Yoo, Shinjae and Cha, Jiook},
  journal={Imaging Neuroscience},
  volume={3},
  pages={imag\_a\_00440},
  year={2025},
  publisher={MIT Press 255 Main Street, 9th Floor, Cambridge, Massachusetts 02142, USA~…}
}

@article{esteban2019fmriprep,
  title={fMRIPrep: a robust preprocessing pipeline for functional MRI},
  author={Esteban, Oscar and Markiewicz, Christopher J and Blair, Ross W and Moodie, Craig A and Isik, A Ilkay and Erramuzpe, Asier and Kent, James D and Goncalves, Mathias and DuPre, Elizabeth and Snyder, Madeleine and others},
  journal={Nature methods},
  volume={16},
  number={1},
  pages={111--116},
  year={2019},
  publisher={Nature Publishing Group US New York}
}

@article{wang2025dca,
  title={DCA: Graph-Guided Deep Embedding Clustering for Brain Atlases},
  author={Wang, Mo and Peng, Kaining and Tang, Jingsheng and Wen, Hongkai and Liu, Quanying},
  journal={arXiv preprint arXiv:2509.01426},
  year={2025}
}

@article{yang2024brainmass,
  title={Brainmass: Advancing brain network analysis for diagnosis with large-scale self-supervised learning},
  author={Yang, Yanwu and Ye, Chenfei and Su, Guinan and Zhang, Ziyao and Chang, Zhikai and Chen, Hairui and Chan, Piu and Yu, Yue and Ma, Ting},
  journal={IEEE Transactions on Medical Imaging},
  year={2024},
  publisher={IEEE}
}

@article{ye2023explainable,
  title={Explainable fMRI-based brain decoding via spatial temporal-pyramid graph convolutional network},
  author={Ye, Ziyuan and Qu, Youzhi and Liang, Zhichao and Wang, Mo and Liu, Quanying},
  journal={Human Brain Mapping},
  volume={44},
  number={7},
  pages={2921--2935},
  year={2023},
  publisher={Wiley Online Library}
}

@article{adhd2012adhd,
  title={The ADHD-200 consortium: a model to advance the translational potential of neuroimaging in clinical neuroscience},
  author={ADHD-200 consortium},
  journal={Frontiers in systems neuroscience},
  volume={6},
  pages={62},
  year={2012},
  publisher={Frontiers Research Foundation}
}

@article{dong2025brain,
  title={Brain Harmony: A Multimodal Foundation Model Unifying Morphology and Function into 1D Tokens},
  author={Dong, Zijian and Li, Ruilin and Chong, Joanna Su Xian and Dehestani, Niousha and Teng, Yinghui and Lin, Yi and Li, Zhizhou and Zhang, Yichi and Xie, Yapei and Ooi, Leon Qi Rong and others},
  journal={arXiv preprint arXiv:2509.24693},
  year={2025}
}

@article{van2013wu,
  title={The WU-Minn human connectome project: an overview},
  author={Van Essen, David C and Smith, Stephen M and Barch, Deanna M and Behrens, Timothy EJ and Yacoub, Essa and Ugurbil, Kamil and Wu-Minn HCP Consortium and others},
  journal={Neuroimage},
  volume={80},
  pages={62--79},
  year={2013},
  publisher={Elsevier}
}

@article{belliveau1991functional,
  title={Functional mapping of the human visual cortex by magnetic resonance imaging},
  author={Belliveau, Jack W and Kennedy, David N and McKinstry, Robert C and Buchbinder, Bradley R and Weisskoff, Robert M and Cohen, Mark S and Vevea, JM and Brady, Thomas J and Rosen, Bruce R},
  journal={Science},
  volume={254},
  number={5032},
  pages={716--719},
  year={1991},
  publisher={American Association for the Advancement of Science}
}

@article{Satterthwaite2016ThePN,
  title={The Philadelphia Neurodevelopmental Cohort: A publicly available resource for the study of normal and abnormal brain development in youth},
  author={Theodore D. Satterthwaite and John J. Connolly and Kosha Ruparel and Monica E. Calkins and Chad Jackson and Mark A. Elliott and David R. Roalf and Karthik Prabhakaran and Ryan Hopson and Meckenzie Behr and Haijun Qiu and Frank D. Mentch and Rosetta Chiavacci and Patrick M. A. Sleiman and Ruben C. Gur and Hakon Hakonarson and Raquel E. Gur},
  journal={NeuroImage},
  year={2016},
}

@article{glasser2013minimal,
  title={The minimal preprocessing pipelines for the Human Connectome Project},
  author={Glasser, Matthew F and Sotiropoulos, Stamatios N and Wilson, J Anthony and Coalson, Timothy S and Fischl, Bruce and Andersson, Jesper L and Xu, Junqian and Jbabdi, Saad and Webster, Matthew and Polimeni, Jonathan R and others},
  journal={Neuroimage},
  volume={80},
  pages={105--124},
  year={2013},
  publisher={Elsevier}
}

@article{marek2011parkinson,
  title={The Parkinson progression marker initiative (PPMI)},
  author={Marek, Kenneth and Jennings, Danna and Lasch, Shirley and Siderowf, Andrew and Tanner, Caroline and Simuni, Tanya and Coffey, Chris and Kieburtz, Karl and Flagg, Emily and Chowdhury, Sohini and others},
  journal={Progress in neurobiology},
  volume={95},
  number={4},
  pages={629--635},
  year={2011},
  publisher={Elsevier}
}

@article{craddock2013neuro,
  title={The neuro bureau preprocessing initiative: open sharing of preprocessed neuroimaging data and derivatives},
  author={Craddock, Cameron and Benhajali, Yassine and Chu, Carlton and Chouinard, Francois and Evans, Alan and Jakab, Andr{\'a}s and Khundrakpam, Budhachandra Singh and Lewis, John David and Li, Qingyang and Milham, Michael and others},
  journal={Frontiers in Neuroinformatics},
  volume={7},
  number={27},
  pages={5},
  year={2013}
}

@article{bellec2017neuro,
  title={The neuro bureau ADHD-200 preprocessed repository},
  author={Bellec, Pierre and Chu, Carlton and Chouinard-Decorte, Francois and Benhajali, Yassine and Margulies, Daniel S and Craddock, R Cameron},
  journal={Neuroimage},
  volume={144},
  pages={275--286},
  year={2017},
  publisher={Elsevier}
}

@article{Beyer2022FlexiViTOM,
  title={FlexiViT: One Model for All Patch Sizes},
  author={Lucas Beyer and Pavel Izmailov and Alexander Kolesnikov and Mathilde Caron and Simon Kornblith and Xiaohua Zhai and Matthias Minderer and Michael Tschannen and Ibrahim M. Alabdulmohsin and Filip Pavetic},
  journal={2023 IEEE/CVF Conference on Computer Vision and Pattern Recognition (CVPR)},
  year={2022},
  pages={14496-14506},
  url={https://api.semanticscholar.org/CorpusID:254685937}
}

@inproceedings{Peng2025WholebrainTR,
  title={Whole-brain Transferable Representations from Large-Scale fMRI Data Improve Task-Evoked Brain Activity Decoding},
  author={Yueh-Po Peng and Vincent K.M. Cheung and Li Su},
  year={2025},
  url={https://api.semanticscholar.org/CorpusID:280391640}
}

@article{Sun2025VoxelLevelBS,
  title={Voxel-Level Brain States Prediction Using Swin Transformer},
  author={Yifei Sun and Daniel Chahine and Qinghao Wen and Tianming Liu and Xiang Li and Yixuan Yuan and Fernando Calamante and Jinglei Lv},
  journal={ArXiv},
  year={2025},
  volume={abs/2506.11455},
  url={https://api.semanticscholar.org/CorpusID:279392087}
}

@article{jack2008alzheimer,
  title={The Alzheimer's disease neuroimaging initiative (ADNI): MRI methods},
  author={Jack Jr, Clifford R and Bernstein, Matt A and Fox, Nick C and Thompson, Paul and Alexander, Gene and Harvey, Danielle and Borowski, Bret and Britson, Paula J and L. Whitwell, Jennifer and Ward, Chadwick and others},
  journal={Journal of Magnetic Resonance Imaging: An Official Journal of the International Society for Magnetic Resonance in Medicine},
  volume={27},
  number={4},
  pages={685--691},
  year={2008},
  publisher={Wiley Online Library}
}

@article{kawahara2017brainnetcnn,
  title={BrainNetCNN: Convolutional neural networks for brain networks; towards predicting neurodevelopment},
  author={Kawahara, Jeremy and Brown, Colin J and Miller, Steven P and Booth, Brian G and Chau, Vann and Grunau, Ruth E and Zwicker, Jill G and Hamarneh, Ghassan},
  journal={NeuroImage},
  volume={146},
  pages={1038--1049},
  year={2017},
  publisher={Elsevier}
}

@article{li2021braingnn,
  title={Braingnn: Interpretable brain graph neural network for fmri analysis},
  author={Li, Xiaoxiao and Zhou, Yuan and Dvornek, Nicha and Zhang, Muhan and Gao, Siyuan and Zhuang, Juntang and Scheinost, Dustin and Staib, Lawrence H and Ventola, Pamela and Duncan, James S},
  journal={Medical Image Analysis},
  volume={74},
  pages={102233},
  year={2021},
  publisher={Elsevier}
}

@inproceedings{qu2024uncovering,
  title={Uncovering cognitive taskonomy through transfer learning in masked autoencoder-based fMRI reconstruction},
  author={Qu, Youzhi and Xia, Junfeng and Jian, Xinyao and Li, Wendu and Peng, Kaining and Liang, Zhichao and Wu, Haiyan and Liu, Quanying},
  booktitle={International Workshop on Human Brain and Artificial Intelligence},
  pages={35--50},
  year={2024},
  organization={Springer}
}

@inproceedings{qu2022transfer,
  title={Transfer learning to decode brain states reflecting the relationship between cognitive tasks},
  author={Qu, Youzhi and Jian, Xinyao and Che, Wenxin and Du, Penghui and Fu, Kai and Liu, Quanying},
  booktitle={International Workshop on Human Brain and Artificial Intelligence},
  pages={110--122},
  year={2022},
  organization={Springer}
}

@inproceedings{zamir2018taskonomy,
  title={Taskonomy: Disentangling task transfer learning},
  author={Zamir, Amir R and Sax, Alexander and Shen, William and Guibas, Leonidas J and Malik, Jitendra and Savarese, Silvio},
  booktitle={Proceedings of the IEEE conference on computer vision and pattern recognition},
  pages={3712--3722},
  year={2018}
}

\clearpage
\begin{center}
{\Large\bfseries Supplementary Material}
\end{center}
\medskip
\renewcommand{\thesection}{\Alph{section}}
\renewcommand{\theHsection}{supp.\Alph{section}}
\renewcommand{\theHsubsection}{supp.\Alph{section}.\arabic{subsection}}
\renewcommand{\theHtable}{supp.\arabic{table}}
\renewcommand{\theHfigure}{supp.\arabic{figure}}
\renewcommand{\theHequation}{supp.\arabic{equation}}
\setcounter{section}{0}

\renewcommand{\theparagraph}{\thesection.\arabic{paragraph}}
\setcounter{paragraph}{0}

\section{Training details}
We trained our model following the settings outlined in Table~\ref{tab:pretraining_settings} for pre-training and downstream evaluation. The data were split into training, validation, and test sets with a ratio of 7:1:2. During the pre-training phase, model selection was based on the validation loss, with the best model was selected at epoch 54 for downstream evaluation. For the downstream evaluation, we conducted a grid search over key hyperparameters, including learning rate, weight decay, and the classifier MLP ratio. This search was performed with a fixed seed, and the optimal hyperparameter configuration for each dataset was determined across three random seeds.
For both pre-training and downstream evaluation, we conducted our experiments on NVIDIA-L40S (48GB) devices with Batch sizes and gradient accumulation steps given below (Table~\ref{tab:pretraining_settings})

\begin{table}[!h]
\centering
\caption{Hyperparameter settings of our model. (GAS: Gradient accumulation steps; BS: Batch size)}
\label{tab:pretraining_settings}
\begin{tabular}{ll}
\toprule
\textit{config} & \textit{value} \\
\midrule
\multicolumn{2}{c}{\textbf{Pre-train configs}} \\
\midrule
optimizer & AdamW  \\
optimizer momentum & $\beta_1, \beta_2 = 0.9, 0.999$ \\
learning rate schedule & warmup cosine schedule \\
learning rate & $5 \times 10^{-4}$ \\
weight decay & 0.05 \\
warmup epochs & 3 \\
total batch size & $4 GAS \times 4$ BS = 16 \\
training epochs & 100 \\
EMA momentum schedule & linear \\
EMA start momentum & 0.996 \\
EMA final momentum & 0.9999 \\
temporal duration & 6s \\
spatial resolution & 12mm \\
encoder depth & 24 \\
predictor depth & 2 \\
mask ratio & 0.65 \\
embedding dimension & 512 \\
\midrule
\multicolumn{2}{c}{\textbf{Downstream configs}} \\
\midrule
learning rate schedule & linear warmup and decay \\
learning rate & $1 \times 10^{-5}$ \\
weight decay & 0.05 \\
total batch size & $4 GAS \times 4$ BS = 16 \\
warmup epochs & 2 \\
training epochs & 30 \\
classifier head depth & 2 \\
classifier head mlp ratio & 4.0 \\
classifier head drop rate & 0.1 \\
\bottomrule
\end{tabular}
\end{table}

\section{Robustness and Additional Analyses}

This section collects the additional analyses added after reviewer discussion. The goal is to separate resolution heterogeneity, which FlexiBrain is designed to handle, from broader scanner-center domain shift, which remains a harder deployment problem for native-space modeling.

\begin{table}[t]
\centering
\caption{Resolution-stratified performance on the held-out PPMI dataset. Values are BACC/wF1 (\%). Label counts are Prodromal/PD.}
\label{tab:supp_ppmi_resolution}
\resizebox{0.72\linewidth}{!}{
\begin{tabular}{lccc}
\toprule
\textbf{Group} & \textbf{n} & \textbf{Labels} & \textbf{BACC / wF1} \\
\midrule
All, TR=2.5s & 90 & 61 / 29 & 69.39 / 74.05 \\
Voxel 3.5$\times$3.5$\times$3.5 & 58 & 36 / 22 & \textbf{73.86 / 74.40} \\
Voxel 3.5$\times$3.5$\times$3.6 & 32 & 25 / 7 & 48.00 / 66.96 \\
\bottomrule
\end{tabular}
}
\end{table}

PPMI is not used during pretraining and therefore serves as a leave-one-dataset-out external test. The lower-resolution bin is smaller and more imbalanced, so we interpret the result cautiously; nevertheless, the stratification shows that the external performance is not explained solely by one averaged metric over a single acquisition setting.

\begin{table}[t]
\centering
\caption{Spatial and temporal downsampling on PPMI. Results are reported as BACC/wF1 (\%).}
\label{tab:supp_downsampling}
\resizebox{0.58\linewidth}{!}{
\begin{tabular}{lcc}
\toprule
\textbf{Condition} & \textbf{Pretraining} & \textbf{End-to-end} \\
\midrule
Original & \textbf{67.05 / 68.97} & \textbf{70.83 / 68.12} \\
T $\times 2$ & 63.38 / 65.52 & 65.52 / 62.55 \\
S $\times 2$ & 61.74 / 65.36 & 67.24 / 67.57 \\
S+T $\times 2$ & 58.96 / 62.55 & 63.79 / 63.62 \\
\bottomrule
\end{tabular}
}
\end{table}

Starting from the original native-resolution PPMI data, we generated degraded variants by reducing temporal resolution, spatial resolution, or both by a factor of two. The performance drop under degradation, especially joint spatiotemporal downsampling, suggests that the model remains sensitive to information carried by higher-resolution fMRI while using physical-unit tokenization as a stable interface for parameter sharing.

\begin{table}[t]
\centering
\caption{Additional gender classification results. Values are ACC/wF1 (\%).}
\label{tab:supp_gender}
\resizebox{0.52\linewidth}{!}{
\begin{tabular}{lcc}
\toprule
\textbf{Model} & \textbf{ABIDE} & \textbf{ADHD-200} \\
\midrule
NeuroSTORM & 80.00 / 75.06 & 59.47 / 54.34 \\
FlexiBrain & \textbf{83.33 / 76.77} & \textbf{71.05 / 70.08} \\
\bottomrule
\end{tabular}
}
\end{table}

We additionally conducted a stricter zero-shot cross-site analysis on ABIDE using leave-one-scanner-center-out evaluation, where the held-out center is unseen during both pretraining and downstream training. Performance degrades in this setting. This result separates resolution heterogeneity from broader scanner-center domain shift: FlexiBrain can process heterogeneous voxel spacings, but transfer to completely unseen acquisition centers may require harmonization, nuisance control, or domain adaptation.

\section{Details of baseline models}

\subsection{Introduction of baseline models} In this section, we introduce our baselines models. First three foundational models for brain activity prediction: BrainMASS~\cite{yang2024brainmass}, Brain-JEPA~\cite{dong2024brain}, and BrainHarmonix~\cite{dong2025brain}, each of which uses distinct parcellation schemes and is trained on large-scale, multi-site fMRI datasets. We also fine-tuned our downstream tasks using other end-to-end models: BrainNetCNN~\cite{kawahara2017brainnetcnn}, BrainGNN~\cite{li2021braingnn} and SwiFT~\cite{kim2023swift}.

BrainMASS utilizes the Schaefer 100-region atlas to parcellate brain networks. It was trained across 26 datasets, totaling 64,584 subjects, including data from UKB, HCP, and ADHD-200. This model focuses on learning representations of functional brain networks by incorporating both masked ROI modeling and latent representation alignment techniques to improve its diagnostic capabilities.

Brain-JEPA integrates a more complex parcellation strategy by combining the Schaefer-400 cortical regions with the Tian-Scale III subcortical regions, resulting in 450 regions of interest (ROIs). Trained on 80 \% of the UKB data, it employs a joint-embedding predictive architecture, incorporating both spatial and temporal masking to capture dynamic brain activity patterns.

BrainHarmonix is a recent multimodal framework designed to harmonize structural and functional neuroimaging signals while addressing variability in acquisition protocols (for example, differences in repetition times in fMRI). The model jointly incorporates structural MRI (T1-weighted) data along with functional connectivity from fMRI, introducing a pretraining strategy that aligns representations across different modalities and scanning conditions.

BrainNetCNN is a convolutional neural network specifically designed for analyzing connectome-structured data. Unlike traditional models that operate on images or graphs, BrainNetCNN uses specialized filters---edge-to-edge, edge-to-node, and node-to-graph convolutional filters---that directly model pairwise connectivity patterns. In our experiments, the model takes as input a functional connectivity (FC) matrix, which is computed using Pearson correlation coefficients between regional fMRI time series.

BrainGNN extends the capabilities of graph neural networks to analyze brain data at the population level by learning interpretable node- and region-specific biomarkers. For each subject, the FC matrix serves as the node-level feature representation, while the target label provides supervision to facilitate the discovery of anatomically meaningful subgraphs. Partial correlations between fMRI time series are utilized as edge attributes, enabling the model to capture conditional dependencies between brain regions rather than just simple pairwise correlations.

SwiFT (Swin 4D fMRI Transformer) tackles the challenge of modeling high-dimensional spatiotemporal brain dynamics by learning directly from raw 4D fMRI volumes, avoiding the information loss of hand-crafted features~\cite{kim2023swift}. Through its 4D windowed attention and efficient architecture, SwiFT outperforms recent state-of-the-art models on large-scale datasets. Additionally, contrastive self-supervised pretraining further enhances its downstream performance, highlighting SwiFT's effectiveness as an end-to-end framework for functional brain imaging.

\subsection{Hyperparameters \& Metric} For a fair comparison of the baseline models, we conduct baseline experiments with three random seeds: {0, 42, 999}, as opposed to the random split policy used in BrainHarmonix~\cite{dong2025brain}. This approach helps us evaluate the stability of different architectures. We perform a grid search for key hyperparameters while also adhering to the default hyperparameter settings when provided. To address the misleading nature of accuracy on imbalanced datasets, we report weighted F1 scores.

The hyperparameter sweeps involve combinations of learning rates ($lr \in \{0.01, 0.005, 0.001, 0.0005, 0.0001\}$) and weight decay ($wd \in \{0.05, 0.005, 0.0005\}$). After completing the hyperparameter sweeps for a fixed seed, we select the best hyperparameter combinations per dataset based on two criteria: either the combination achieves the highest accuracy, or it is the closest to the default hyperparameters.

For the multi-modality model BrainHarmonix~\cite{dong2025brain}, we follow the guidelines provided in the official code. We conduct T1 and fMRI fusion experiments on the ADNI and PPMI datasets, as well as fMRI-based experiments on all available datasets. Due to a lack of detailed information, we report the results of the Full Finetuned BrainHarmonix-F model, following a similar process to that of BrainHarmonix. The baseline models are finetuned on a slurm system with NVIDIA-L40S (48GB) devices. BrainHarmonixs are trained using the distributed training to accelerate the process. The batch size may vary to fully utilize the GPU memory.

\subsection{Ablation on data size}

We further evaluate the data efficiency of our model under a few-shot fine-tuning setting on the PPMI dataset. The data are partitioned into training, validation and test sets with a 7:1:2 ratio. We then subsample the training split and fine-tune the model using only 10 \%, 20 \%, 40 \%, 80 \% and 100 \% of the available training samples for the three-class classification task. As reported in Table~\ref{tab:data_portion}, our model achieved 58.33 \% ACC and 49.92 \% F1 when only 10 \% of the training data are used, and its performance improves steadily as more data become available, reaching 76.04 \% ACC and 71.94 \% F1 with the full training set. These results highlight the strong label efficiency and robustness of our approach in low-data regimes.

\begin{table}[!h]
\centering
\caption{Few-shot performance on PPMI}
\resizebox{0.6\textwidth}{!}{
\begin{tabular}{lccccc}
\toprule
\textbf{Data Portion} & 10\% & 20\% & 40\% & 80\% & 100\% \\
\midrule
ACC \% & 58.33 & 62.50 & 63.54 & 66,67 & 76.04 \\
F1 \% & 49.92 & 58.63 & 61.66 & 64.42 & 71.94 \\
\bottomrule
\end{tabular}
}
\label{tab:data_portion}
\end{table}

\subsection{Ablation on patch size}
We systematically ablate the temporal duration $\tau$ and the spatial resolution $\rho$ as shown in main text. We report the detailed values in Table~\ref{tab:ablation:patch}. We evaluated performance with ADHD by training it from scratch and find the best settings with $\tau = 6 $ seconds and $\rho = 12$ millimeters.  Increasing $\tau$ beyond 6 seconds and $\rho$ beyond 12 millimeters leads to coarser patches that blur details, while very small $\tau $or voxel spacings cause discretization, reducing sensitivity to subtle differences and harming generalization across protocols.

\begin{table}[!h]
\centering
\caption{Ablation study of patch size on the ADHD dataset.}
\resizebox{0.7\textwidth}{!}{
\begin{tabular}{l>{\columncolor{rowgray}}c>{\columncolor{rowgray}}ccccc}
\toprule
\multirow{2}{*}{\diagbox{\textbf{Space}}{\textbf{Temporal}}}
& \multicolumn{2}{c}{6s}
& \multicolumn{2}{c}{12s}
& \multicolumn{2}{c}{18s}\\
\cmidrule(lr){2-3}\cmidrule(lr){4-5}\cmidrule(lr){6-7}
   & ACC\% & F1  \% & ACC \% & F1  \%   & ACC\% & F1  \% \\
\midrule
9mm    & 64.95 & 57.07 & 62.89 & 55.62 & 61.86 & 58.76 \\
\rowcolor{rowgray}
12mm   & \textbf{65.98} & 61.37 & 64.95 & 57.07 & 64.95 & \textbf{64.95} \\
18mm   & 63.92 & 57.34 & 61.86 & 60.65 & 61.86 & 58.76 \\
24mm   & 61.86 & 54.90 & 60.82 & 50.76 & 59.79 & 55.88 \\
\bottomrule
\end{tabular}
}
\label{tab:ablation:patch}
\end{table}

\subsection{Analysis of JEPA Collapse}
\begin{figure*}[t]
  \centering
   \includegraphics[width=\linewidth]{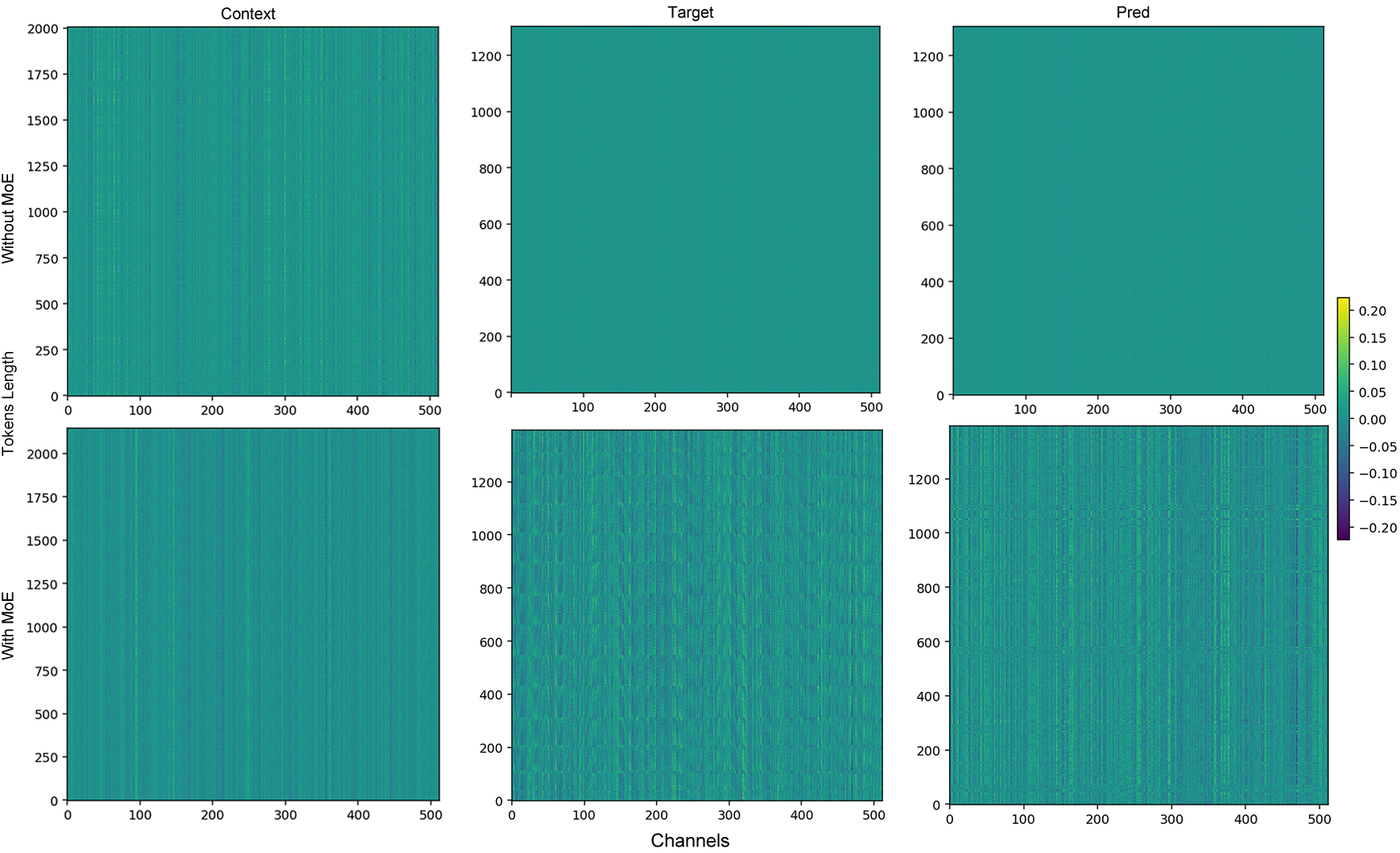}
   \caption{\textbf{Feature heatmaps} showing the outputs of the context encoder, target encoder and predictor. The features in the upper section, without MoE, exhibit simpler patterns with lower variability in both the features and predictions. In contrast, the features in the lower section, with MoE, demonstrate a more complex representation of the context and target features, along with more diverse and accurate predictor outputs.
   }
   \label{fig:fea_coll}
\end{figure*}

To systematically analyze the collapse of JEPA more meticulously, we computed the mean Effective Rank and Cosine Similarity of features as shown in the Table~\ref{tab:metrics_coll}, in the model with MoE, the context features exhibit a higher effective rank and lower cosine similarity, indicating greater feature independence and diversity. In contrast, the prediction features in the model without MoE have a lower effective rank (60.15) and higher cosine similarity (0.7967), suggesting greater redundancy and potential information compression. The target features in the MoE model are also more dispersed, reflecting a broader distribution. To qualitatively analyze the collapse, we visualize the features output by the two encoders and the predictor. It is evident that the features from the model with MoE exhibit greater detail and dynamics, rather than converging into a limited set of patterns. (See Fig.~\ref{fig:fea_coll}).

\begin{table}[!h]
\centering
\caption{  Collapse metrics }
\resizebox{0.6\textwidth}{!}{
\begin{tabular}{lccc}
\toprule
\multirow{2}{*}{}  &
\textbf{MoE} & \textbf{Effective Rank $\uparrow$ } & \textbf{Cosine Similarity $\downarrow$} \\
\midrule
Context   & \xmark & 145.94  & 0.4641  \\
Target  & \xmark & 60.15 & 0.7967 \\
Pred  & \xmark & 207.66  & 0.6786  \\
\rowcolor{rowgray}
Context & \cmark & 237.18  & 0.4181  \\
\rowcolor{rowgray}
Target & \cmark & 144.27  & 0.3247  \\
\rowcolor{rowgray}
Pred  & \cmark & 178.37  & 0.2274  \\
\bottomrule
\end{tabular}
}
\label{tab:metrics_coll}
\end{table}

\begin{figure}[t]
  \centering
   \includegraphics[width=\linewidth]{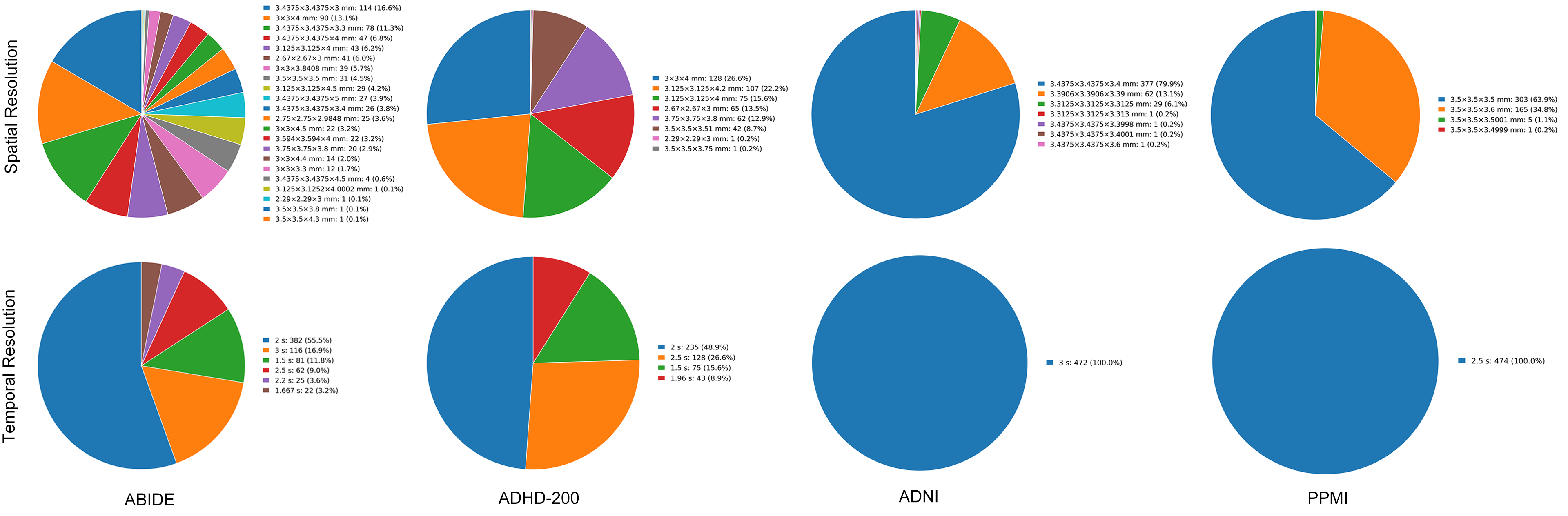}
   \caption{\textbf{Distribution of Spatiotemporal Resolution}
   }
   \label{fig:resolution}
\end{figure}

\section{Details of datasets and task}

In this section, we detailed all the datasets with classification task. And we also give the distribution of spatiotemporal resolutions of different datasets (see Fig.~\ref{fig:resolution}).

\subsection{Autism Brain Imaging Data Exchange}
The Autism Brain Imaging Data Exchange (ABIDE) is a multi-site, open-access initiative that aggregates structural and resting-state fMRI scans - alongside rich phenotypic data - from individuals with autism spectrum disorder (ASD) and matched typically developing controls to accelerate reproducible neuroimaging research. In our experiments, we utilize 688 subjects~\cite{di2014autism}.

\subsection{Attention Deficit Hyperactivity Disorder}
The Attention Deficit Hyperactivity Disorder (ADHD-200) is an open multi-site neuroimaging resource with resting-state fMRI across eight contributing centers. For the classification task we merged all ADHD subtypes (Combined/Inattentive/Hyperactive-Impulsive) into a single ADHD label and treat typically developing controls as TDC with 481 subjects~\cite{adhd2012adhd}.

\subsection{Alzheimer's Disease Neuroimaging Initiative}
The Alzheimer's Disease Neuroimaging Initiative (ADNI) is a large-scale, longitudinal, multi-center project initiated in 2004, designed to collect and publicly share rich multimodal biomarker, imaging, genetic, biofluid and clinical data from cognitively normal older adults, individuals with mild cognitive impairment (MCI) and patients with Alzheimer's disease (AD).
For our task, we separated disorders into two classification with healthy controls at a total of 239 subjects~\cite{jack2008alzheimer}.

\subsection{Parkinson's Progression Markers Initiative}
The Parkinson's Progression Markers Initiative (PPMI) is an ongoing, international, longitudinal observational study launched in 2010 by The Michael J. Fox Foundation to identify and validate biomarkers of Parkinson's disease (PD) risk, onset, and progression.
We combined two different diseases and healthy controls from the PPMI dataset to create a three-class classification task with 474 subjects~\cite{marek2011parkinson}.

\section{Details of Transfer learning}
We use transfer learning to reveal the domain relationship for interpretability.
Let $(x_T, y_T)$ denote target samples and labels, and let $\theta_S$ be the parameters pretrained on the source domain.
Transfer learning adapts the source model to the target by optimizing
\begin{equation}
\min_{\theta_T}\
\mathbb{E}_{(x_T, y_T)}
\left[
\mathcal{L}\big(f(x_T;\theta_T \!\mid\! \theta_S),\, y_T\big)
\right],
\end{equation}
where $f(\cdot;\theta_T \!\mid\! \theta_S)$ represents the target model initialized from $\theta_S$,
and $\mathcal{L}$ is loss on the target domain.
In practice, the encoder from $\theta_S$ is frozen, and only the lightweight decoder is fine-tuned on a small subset of target data to evaluate cross-task transferability. This design ensures that differences in transfer performance primarily reflect the quality and domain-specificity of the learned representations, rather than optimization effects or capacity differences. All transfers share identical training schedules, batch sizes, and optimization hyperparameters, following the same settings as pretraining where applicable, to ensure experimental consistency.
The transferability from a source $s$ to a target $t$ is quantified by the \emph{transfer gap}:
\begin{equation}
\Delta L_{s\to t} = L_t^{\mathrm{gold}} - L_{s\to t},
\end{equation}
where $L_{s\to t}$ is the target loss after transfer learning and $L_t^{\mathrm{gold}}$ represents the loss of the gold model trained on the full target domain.
A negative $\Delta L_{s\to t}$ indicates positive transfer, suggesting that knowledge from the source domain improves the target performance. This framework allows us to systematically quantify domain relationships, assess the robustness of representations under domain shift, and reveal how diagnostic categories (AD, MCI, CN) influence the ease of cross-domain adaptation.


\end{document}